Local transient rheological behavior of concentrated suspensions

F. Blanc, F. Peters and E. Lemaire
CNRS - Université de Nice
LPMC - UMR 6622
Parc Valrose 06108 Nice cedex 2 – France

**Synopsis**

This paper reports experiments on the shear transient response of concentrated non-Brownian suspensions. The shear viscosity of the suspensions is measured using a wide-gap Couette rheometer equipped with a Particle Image Velocimetry (PIV) device that allows measuring the velocity field. The suspensions made of PMMA particles (31µm in diameter) suspended in a Newtonian index- and density-matched liquid are transparent enough to allow an accurate measurement of the local velocity for particle concentrations as high as 50%. In the wide-gap Couette cell, the shear induced particle migration is evidenced by the measurement of the time evolution of the flow profile. A peculiar radial zone in the gap is identified where the viscosity remains constant. At this special location, the local particle volume fraction is taken to be the mean particle concentration. The local shear transient response of the suspensions when the shear flow is reversed is measured at this point where the particle volume fraction is well defined. The local rheological measurements presented here confirm the macroscopic measurements of Gadala-Maria and Acrivos (1980). After shear reversal, the viscosity undergoes a step-like reduction, decreases slower and passes through a minimum before increasing again to reach a plateau. Upon varying the particle concentration, we have been able to show that the minimum and the plateau viscosities do not obey the same scaling law with respect to the particle volume fraction. These experimental results are consistent with the scaling predicted by Mills and Snabre (2009) and with the results of numerical simulation performed on random suspensions [Sierou and Brady (2001)]. The minimum seems to be associated with the viscosity of an isotropic suspension, or at least of a suspension whose particles do not interact through non-hydrodynamic forces, while the plateau value would correspond to the viscosity of a suspension structured by the shear where the non-hydrodynamic forces play a crucial role.

**I. INTRODUCTION**

Concentrated suspensions are very common in several engineering fields such as civil engineering, food or pharmaceutical industry or geophysical situations such as debris flows, sediment transport, and submarine avalanches. In order to understand their flowing behavior, it is desirable to know their response to imposed forces and motions at their boundary. A broad variety of complex rheological behaviors such as shear thinning or shear thickening, yield stress, normal stress differences or shear banding are exhibited. This diversity issues from the diversity of the interactions (Brownian, Van der Waals, steric, electrostatic…) and of the particles properties (size, shape, polydispersity…) that are possibly involved in the flow behavior. Nevertheless even a suspension whose physicochemical characteristics are very simple (non-Brownian spherical particles, negligible colloidal interactions, Newtonian suspending liquid) can exhibit a complex rheological behavior. For instance, non-Brownian concentrated suspensions are usually shear thinning [Stickel and Powell (2005)] and, if the particle and fluid density are not carefully matched, a yield stress has been shown to develop [Fall et al. (2009)]. The transient response of such suspensions is also of interest as reported by Gadala-Maria and Acrivos (1980), Kolli et al. (2002) and Narumi et al. (2002). As emphasized by Stickel and Powell (2005) and Morris (2009) in their review articles on the rheology of dense suspensions, such a complexity of the rheological behavior is most likely due to the formation of a microstructure induced by the shear flow together with the non-hydrodynamic forces and the key problem is to understand the relationship between the macroscopic or bulk properties of the medium and its microscopic structure.

The first direct experimental evidence of the shear-induced anisotropic microstructure has been provided by Parsi and Gadala-Maria (1987) who measured the pair distribution function for a suspension whose particle volume fraction was 0.4. They note a maximum of the pair distribution function in a direction which is roughly that of the compression axis. Furthermore, they show that when the direction of the shear is reversed, the particles rearrange into the mirror image of the structure. This direct observation of the microstructure confirms the hypothesis advanced by Gadala-Maria and Acrivos (1980), a few years before, to explain the transient shear stress response of a suspension when it undergoes a shear reversal. The suspension viscosity was shown to depend only on the strain after the shear reversal. In 2002, these results were supplemented by the measurement of the transient normal stresses [Kolli et al. (2002), Narumi et al. (2002)] which present the same kind of behavior, i.e. a collapse of the recovery normal stresses if plotted against the strain, whatever the shear rate step magnitude.

The characterization of the microstructure has also motivated a lot of numerical studies based on various numerical techniques such as Stokesian Dynamics [Singh and Nott (2000), Sierou and Brady (2002), Drazer et al. (2004), Bricker and Butler (2007)] or Force Coupling Method [Yeo and Maxey (2010), Abbas et al. (2007)] which all conclude to the formation of an anisotropic shear-induced microstructure.

Achieving accurate rheological measurements with a non colloidal suspension is quite challenging because perturbative effects such as wall slip [Jana et al. (1995)], shear localization [Huang et al. (2005)] or particle migration [Leighton and Acrivos (1987), Chow et al. (1994)] can arise during the experiment. The local rheological measurements have been shown to be very powerful to measure the true rheological response of a dense suspension.

For instance, in 2002 Shapley et al. use Laser Doppler Velocimetry to measure particle velocity fluctuations and velocity profiles in concentrated suspensions sheared in a narrow-gap Couette cell. Another powerful technique to visualize both the velocity and the concentration profiles in suspensions is the Magnetic Resonance Imaging. In 2006, Ovarlez and his collaborators used such MRI techniques to perform local measurements on very concentrated suspensions sheared in a wide-gap Couette viscosimeter. They have been able to measure both the velocity and the particle concentration profiles in the suspension sheared between two concentric cylinders. They have obtained very impressive results on the stationary behavior of suspensions with particle concentration as high as 60%. One of their main results is that contrarily to what is observed with macroscopic rheological measurements, i.e. a pseudo yield stress, suspensions have a purely viscous behavior. Furthermore, they measured the particle concentration profiles and showed that for such high concentrated suspensions, the migration is almost instantaneous. It should be stressed that all their measurements have been performed after a period of pre-shear.

On the opposite, the present paper is focused on the transient shear rate response of rather concentrated suspensions ($0.3<\phi<0.5$) of non-Brownian spheres (diameter 31µm) when a shear stress is suddenly applied or reversed. Suspensions of various concentrations are sheared between two concentric cylinders so that the shear stress field is controlled in the gap. The suspensions are designed to be almost transparent allowing to determine the velocity field in the gap using PIV measurements. The local transient response of the suspension is then deduced from the local shear stress and shear rate. The main characteristics of the suspensions are presented in section II. The third section is devoted to the description of the experimental procedure. The results are displayed in section IV where two transient regimes with two different characteristic times are studied. The long time variation of the velocity profile coming from the shear induced particle migration is presented in Sec. IV.A while the rapid changes happening after the shear reversal are described in Sec. IV.B. All these results are discussed in Section V.

**II. SUSPENSION CHARACTERISTICS**

The suspension is made of monodisperse spherical PMMA particles (Microbeads CA30), 31µm in diameter, dispersed in a Newtonian fluid that has been specially designed for our application (Cargille Laboratories, Immersion Liquid Code 11295101160). The fluid has both the same density (d=1.18) and the same refractive index (n=1.49) as the PMMA particles in order to perform the PIV measurements detailed below. To improve the index matching, the temperature is controlled and the best matching is obtained for T=34°C. At this temperature, the suspending fluid has been found to be fully Newtonian and its viscosity has been measured to be $\eta_0$=1.02 Pa.s. [Blanc et al. (2011)]. The particle volume fraction varies from 0.3 to 0.5. The shear viscosity measurements have been performed for shear rates, $\dot{\gamma}$, between $10^{-2}$ and 1 $s^{-1}$. Therefore, given the above characteristics of the suspension, the Peclet number that measures the ratio of convection to thermal diffusion is very large:

$$\text{Pe} = \frac{6\pi\eta_0 a^3 \dot{\gamma}}{kT} \sim 10^5 - 10^7 \qquad .$$

and the Reynolds number based on the particle size is very small:

$$\text{Re}_p = \frac{\rho \dot{\gamma} a^2}{\eta_0} \sim 10^{-9} - 10^{-7}$$

where ρ is the density of the suspending fluid and a, the radius of the suspended particles. Thus, our measurements were carried out in the creeping flow regime with negligible Brownian motion.

The viscosity of the suspensions has been studied either in a narrow- or in a wide-gap Couette geometry mounted on a rheometer Mars II (Haake, Thermofisher). The possible variation of the viscosity of the suspension with the shear rate was first studied in a narrow-gap Couette cell (inner radius, 10 mm, outer radius, 10.85 mm). After the suspension had been poured in the cup and degassed, the shear stress was increased step by step. At each step, the steady value of the shear rate was recorded and the value of the viscosity deduced. The measurements have then been performed upon increasing or decreasing the stress, and gave the same results, suggesting that no migration had occurred. Figure 1 shows the flow curve obtained for a $\phi=0.47$ suspension. The data are well fitted by a power law: $\sigma = 40\dot{\gamma}^{0.88}$ where $\sigma$ and $\dot{\gamma}$ are measured in Pa and $s^{-1}$, respectively. Such a slight shear thinning behavior is often observed in concentrated non-Brownian suspensions [Gadala-Maria and Acrivos (1980), Zarraga et al. (2000)]. Acrivos and coworkers [Acrivos et al. (1994)] have shown that such a shear thinning could occur when the host liquid and the particles do not have the same density, due to the viscous resuspension of the particles. Nevertheless we never noticed any evidence for sedimentation in our experiments and the origin of the low shear thinning observed in our experiments remains unclear.

As it is well known [Jana et al. (1995)] and shown again in Appendix B, the viscosity measurements are difficult to perform in narrow-gap Couette cell because of the wall slip that significantly changes the value of the apparent viscosity. For that reason and since the observed shear thinning is rather weak (when the shear stress is multiplied by two, the viscosity decreases by only 9%), we decided to perform the further viscosity measurements in the wide-gap Couette cell that will be used for the PIV experiments even though the shear stress is not constant in the gap.

The inner and outer cylinder radii are $R_{in}$=14 mm and $R_{out}$=24 mm respectively, their height is 60 mm. The rotor basis is hollowed out and is positioned one millimeter above the cup (Fig. 2). We have chosen to set the rotor close to the cup in order to limit shear-induced migration from or toward the bottom [Leighton and Acrivos (1987)]. Nevertheless, since the rotor is close to the cup and its immersed height is not very large compared to the gap between the coaxial cylinders, the usual relation between the torque and the shear stress does not hold and a correction has to be introduced:

$$\Gamma = 2\pi r^2 L \cdot \sigma(r) + \Gamma^{bot} \quad (1)$$

Upon varying the height L of the sheared liquid, we measure $\Gamma^{bot}$, the torque that results from the presence of the bottom [Blanc et al. (2011)] and we deduce the shear stress field in the gap between the coaxial cylinders:

$$\sigma(r) = \frac{\Gamma - \Gamma^{bot}}{2\pi L r^2} = \frac{\Gamma^{cor}}{2\pi L r^2} \quad (2)$$

Neglecting the slight shear thinning behavior of the suspension, the shear rate in the gap is obtained from the spin rate of the inner cylinder, $\Omega$:

$$\dot{\gamma}(r) = 2\Omega \frac{R_{in}^2 R_{out}^2}{R_{out}^2 - R_{in}^2} \frac{1}{r^2} \quad (3)$$

And the viscosity is given by the ratio of the shear stress to the shear rate:

$$\eta_{app} = \frac{\Gamma^{cor}}{4\pi \Omega L} \frac{R_{in}^2 - R_{out}^2}{R_{in}^2 R_{out}^2} \quad (4)$$

The variation of the apparent viscosity as a function of the particle volume fraction is shown on Fig. 3. These measurements have been performed for an average shear rate, $\langle \dot{\gamma} \rangle = (\dot{\gamma}(R_{in}) + \dot{\gamma}(R_{out}))/2$, of about $0.02 s^{-1}$.

The data represented in Fig. 3 are fit to a Krieger-Dougherty law (solid line) with a maximum packing fraction $\phi^*=0.536$ and an exponent n=2:

$$\eta = \eta_0 \frac{1}{\left(1 - \frac{\phi}{\phi^*}\right)^n} \quad (5)$$

This value of ϕ* is very small compared to the values usually found in the literature [Ovarlez et al.(2006), Stickel and Powell (2005)] that are around 0.60-0.63. It should be mentioned that Zarraga et al. (2000) have also found a rather low value for the packing fraction (ϕ*=0.58). As discussed in Appendix A, we attribute this low value to polymeric forces between particles. Indeed, we noticed a viscoelastic behavior of the suspensions which is not expected for a suspension of truly hard, non-Brownian spherical particles.

**III. EXPERIMENTAL PROCEDURE**

PIV is a now widespread non-invasive technique that allows for flow profile measurements (see [Adrian (2005)] for a review). However, even if it has been already used in the field of concentrated suspensions [Lenoble et al. (2005)], [Wiederseiner et al. (2009)], PIV is still a challenging technique in such heterogeneous media. The experimental facility and methods are described in detail elsewhere [Blanc et al. (2011)]. The present section summarizes the main information.

A transparent wide-gap Couette cell is mounted on a controlled-stress rheometer (Haake Mars II, Thermo Scientific). The measurement of the rotor spin rate provides macroscopic rheometric data while the velocity profile in the gap between the cylinders is deduced from PIV analysis. A schematic drawing of the set-up is shown on Fig.2. The inner rotating cylinder (28 mm in diameter, 60 mm in height) and the outer stationary cup (48 mm in diameter, 60 mm in height) are made of PMMA, in order that their refractive index is approximately matched with the index of the suspension. The front face has been polished in the form of a rectangular window that lets the horizontal laser sheet (Lasiris TEC Laser 635 nm, 35mW, Stockeryale) enter the gap between the cylinders. The inner cylindrical walls have been roughened to minimize the suspension slip. The whole apparatus is placed in a thermostated box, and the temperature is set to T=34°C.

A small quantity of the particles, that amounts to 0.25% of the whole suspension in volume, is colored with a fluorescent dye, the Nile Blue A (excitation peak at 635 nm, broad emission peak at 650 nm). Those particles serve as flow tracers. A CCD USB camera (Pixelink PLB 741 U, 1280x1024 pix$^2$) is placed under the cup and records sequential images of the illuminated part of the suspension through the polished bottom of the cup. The laser sheet and the camera are trigged by an external oscillator. In order to enhance the contrast of the images, the reflected and scattered light is filtered out by an optical high pass filter ($\lambda_{HP}$=650nm) placed in front of the camera, so that only the fluorescing light is detected. The laser sheet illuminates the horizontal plane 14 mm above the bottom of the cup, where it has been shown [Blanc et al. (2011)] that the influence of the bottom on the velocity profile could be neglected. Fig.4 displays a typical image of the illuminated plane obtained with a $\phi$=44.4% suspension.

The image processing is performed using an open source software DPIV Soft available on the web (https://www.irphe.fr/~meunier/). Each image is divided into overlapping subsets named correlation windows (64x64 pixels). The cross correlation of the corresponding windows from two successive images yields the mean velocity of the particles in the window. The same procedure performed on all windows gives the velocity field (60x30 2D vectors) in the illuminated plane. The concentration of colored particles has been chosen such that the number of particles in a window is sufficient (approximately 5), while keeping acceptable light absorption. Given the size of the image plane (approximately 1cm x 1.5cm), the spatial resolution is around 200 μm. The camera acquires frames at a maximum rate of 10Hz, resulting in a temporal resolution of 100ms.

The radial and azimuthal components of the velocity are then computed, and averaged over the azimuthal angle $\theta$. The mean radial velocity $v_r(r,t)$ is always very close to zero ($v_r/v_\theta \sim O(10^{-2})$). The local shear rate is extracted from the mean azimuthal velocity $v_\theta(r,t)$: the best fit of the function $f(r,t) = a(t)\, r + \dfrac{b(t)}{r} + \dfrac{c(t)}{r^2}$ to the velocity data is sought and the shear rate is deduced from the well-known expression:

$$\dot{\gamma} = r \frac{\partial}{\partial r}\left(\frac{f(r,t)}{r}\right) \qquad (6)$$

As explained previously, the local shear stress is deduced from the torque on the rotating cylinder (eq.(2)) that is corrected for the influence of the bottom of the cup [Blanc et al. (2011)] and the viscosity profile is computed:

$$\eta(r,t) = \frac{\sigma(r,t)}{\dot{\gamma}(r,t)} \qquad (7)$$

The statistical error in the viscosity measurement has been estimated lower than 4% [Blanc et al. (2011)] and the uncertainty in the determination of the bulk volume fraction is less than 0.1%.

## IV. TRANSIENT RESPONSE

### A. long time migration

The main goal of this paper is the measurement of the local transient response of a concentrated suspension to a shear reversal. One parameter plays a major role in this mechanical response, namely the particle volume fraction. As a consequence, we need to efficiently control this parameter. However, it is well known that the particles in a sheared suspension tend to migrate toward regions where the shear rate is lower [Phillips et al. (1992)]. This migration is all the more rapid as the particle concentration is higher and can have significant effects after even a few revolutions of the bob. For example, the velocity profile change obtained for a continuously sheared suspension whose mean volume fraction is $\phi_{mean}$=0.444 is displayed in Fig.5. This curve has been obtained upon controlling the rotor spin rate at a value 1 rpm which corresponds to an average shear rate of $2s^{-1}$. Figure 5 suggests that wall slip is present near the inner cylinder but, due to the finite size of the correlation windows, we are not able to measure accurately the velocity close to the boundaries. So, we cannot conclude anything about the wall slip velocity. After ten revolutions, the velocity profile has changed significantly due to outward migration. From eq. (2), (6) and (7), we can compute the time evolution of the viscosity profile that is shown in Fig.6. We note that around a special value of $r=r_c$, the viscosity remains approximately constant. We have chosen $r_c/R_{out}$ =0.77, which is the position where the viscosity variation is the lowest at the beginning of the migration, even though in Fig.6, where very long time variation are displayed, a higher value seems more appropriate. Since the viscosity mostly depends on the particle volume fraction through a monotonic increasing function, the volume fraction around the position $r=r_c$ should also be constant. The existence of this radial zone where the particle concentration varies slightly throughout the migration process was observed experimentally by Phillips et al. in 1992 and is predicted either by the suspension balance model [Morris and Boulay (1999)] or by the diffusive flux model [Phillips et al. (1992)]. We have observed that the value of $r_c$ depends only very weakly on the mean volume fraction. As a consequence, in the following, all local measurements are performed at this particular radial position in the gap, giving the properties of the suspension at the volume fraction $\phi_{mean}$.

**B. Transient response after shear reversal**

Besides this long time evolution of the velocity profiles, we have observed a fast transient response of the suspension when the shear is reversed. Figure 7 shows the evolution of the normalized velocity profiles in a suspension with $\phi$=44.4% when the direction of the torque applied on the inner cylinder is reversed. The measurement has been performed right after that the suspension had been mixed and poured in the Couette cell, so that an approximately homogeneous spatial distribution of the particles was expected at the beginning of the experiment. A negative torque of 250 µNm is applied to the inner cylinder until it has rotated quarter turn which corresponds to a mean deformation of 3.5 in the suspension. At time zero, the magnitude of the torque passes from -250 µNm to 250 µNm which, according to expression (2) corresponds to a shear stress of 2 Pa at the bob. In Fig. 7, the dashed line represents the velocity profile expected for a Newtonian fluid. The solid lines are the measured profiles at various times, t, after the torque reversal. The very first ones are nearly Newtonian, the next profiles deviate from this Newtonian curve before returning to it for longer times. From each curve, following the procedure described in section III, we can deduce the viscosity profiles ($\eta(r,t)$) as well as the total strain ($\gamma(r,t)$):

$$\gamma(r,t) = \int_0^t \dot{\gamma}(r,t')dt' \qquad (8)$$

In particular, we can measure the viscosity and the strain at the special point r=$r_c$ where we are assured that the particle volume fraction has almost not changed.

If we plot the viscosity $\eta(r_c,t)$ as a function of the strain, $\gamma(r_c,t)$, we obtain the curve that is shown on Fig. 8. Just after the shear reversal, the viscosity undergoes a step-like decrease. When the strain increases further, the viscosity first decreases, passes through a minimum and increases again to reach a plateau value.

First, we can note that such a strain-dependent viscosity is consistent with the time evolution of the normalised velocity profile shown in Fig. 7 since, in a wide-gap Couette cell, the shear rate is larger close to the bob and so is the strain. Thus as long as the plateau viscosity has not been reached everywhere in the gap, the viscosity is not uniform and the velocity profile is not Newtonian.

Second, it should be stressed that the characteristic total strain for such an experiment is 3.5 which corresponds to approximately 1/4 revolution of the bob. According to Fig. 6, for such a strain, the effect of the particle migration on the viscosity profile is negligible. All further experiments were conducted for an accumulated strain lower than 50 (corresponding to 4 revolutions of the bob) so that the particle migration was weak. Moreover, to minimize further the effect of migration, all viscosity measurements were performed at the radial position $r=r_c$.

The shear reversal experiment is repeated for other values of the applied torque corresponding to shear stresses of 0.85, 2.12, 4.25 and 5.95 Pa and the results obtained with a $\phi=0.47$ suspension are presented on Fig. 9. The first observation is that the viscosity mainly depends on the strain amplitude even though, in accordance with the slight shear thinning behavior of the suspension (Fig.1), the value of the plateau decreases as the shear rate increases. On the contrary, the value of the minimum of the viscosity seems not to depend on the value of the applied torque.

The same experiment is carried out for other values of the particle concentration and the results are displayed on Fig. 10. For each particle volume fraction, the value of the applied torque is chosen in order that the angular velocity of the internal cylinder at the plateau is roughly the same (around 0.5 rpm which corresponds to an average shear rate of $0.1s^{-1}$). As the particle concentration increases, the minimum is more pronounced and the strain necessary to reach the plateau value of the viscosity decreases.

A striking result shown on Fig. 11 is that the minimum of the viscosity scales as $\left(\phi_m^* - \phi\right)^{-1}$ with $\phi_m^* = 0.537$ while the plateau viscosity scales as $\left(\phi_p^* - \phi\right)^{-2}$ with $\phi_p^* = 0.534$. This result is discussed in the following section.

## V. DISCUSSION.

The transient response of a concentrated suspension after shear reversal has already been studied through macroscopic rheological measurements either in narrow-gap Couette [Gadala-Maria and Acrivos (1980)] or in parallel plate geometry [Kolli et al. (2002), Narumi et al. (2002)] . In particular, it has been shown that the transient viscosity is a function of the strain only. Gadala-Maria and Acrivos (1980) proposed to explain the transient response by the development of a shear-induced microstructure. This microstructure has been observed by Parsi and Gadala-Maria (1987) and is characterized by the asymmetry of the pair distribution function with an excess of particles in the compressional quadrant. When the direction of the shear is reversed, the microstructure is broken and the particles rearrange into the mirror image of this microstructure.

In 2002, Kolli et al. studied the transient shear and normal force response to a shear reversal in non-Brownian suspensions in an annular plate-plate geometry. They observed that the normal force exhibited a transient behavior too. It abruptly went to a negative small value before increasing again to its positive value.  They connected the initial value of both shear and normal stresses with the hydrodynamic contribution (in a structured state) and the plateau value with a mixed contribution of hydrodynamic and contact forces. They noted that the hydrodynamic contribution compared to the plateau value decreased as the concentration increased. Haan and Steif (1998) found previously the same type of behavior for the particle pressure in numerical simulations of 2D suspensions of cylinders.

Thus both shear-induced microstructure and relative contributions of the hydrodynamic and non-hydrodynamic effects are presumably involved in the transient rheological response of non-Brownian suspensions.

Basically, the local response of the suspension presented in the current paper displays the same features as those obtained through conventional viscosimetric measurements. The main contribution of our work is to confirm that the transient response is actually a bulk property of the flowing suspension and cannot originate from a boundary problem such as the development of a Vand zone [Vand (1948)] or of a slip layer [Jana (1995)].

A more precise comparison between our results and those obtained by Gadala-Maria and Acrivos reveals a small difference: while all their results collapse into a single curve when the viscosity divided by its plateau value is plotted versus the strain, our results are slightly different. Indeed, Figure 9 shows that, as the applied torque increases, the plateau viscosity decreases while the minimum remains almost unchanged. This effect is rather weak so we do not want to give too much importance to this observation but it would mean that the shear thinning behavior of the suspension is related to a change of its microstructure together with the non-hydrodynamic interactions with the magnitude of the shear rate. Indeed, we speculate that the viscosity passes through the minimum when the microstructure has disappeared and the suspension is almost isotropic before the particles have rearranged into the steady-state structure when the viscosity plateau is reached again. In this latter state, the particles are very close together along the compression axis and it is plausible that their arrangement and the non-hydrodynamic interactions depend slightly on the value of the shear rate that causes the plateau viscosity to vary. On the opposite, if the minimum viscosity is reached when the suspension is unstructured and when non-hydrodynamic forces have relaxed, the value of the minimum should not depend on the shear rate magnitude. The results reported in Fig. 11 are more pronounced and have more implications. The minimum viscosity and the plateau value do not scale in the same way with the particle concentration. Both diverge when the particle concentration approaches approximately the same value, $\phi^*=0.535$ but with different power laws of $(\phi^*-\phi)$. The minimum viscosity diverges when $\phi$ tends to $\phi^*$ with an exponent (-1) while the plateau viscosity follows a (-2) power law. Following our speculation that the suspension is almost isotropic when the viscosity is minimum and that an anisotropic structure has formed at the plateau, two different behaviors were indeed expected. As for the plateau, the (-2) power law is usually observed both in experiments and in numerical simulations. From a theoretical point of view, Mills and Snabre (2009) have proposed to explain this scaling with a model that accounts for the solid friction between particles in dynamic clusters. In the same paper, they also provide a scaling of the viscosity with the solid volume fraction for an isotropic random suspension where the dissipation is purely hydrodynamic. The viscosity is shown to scale as $\phi^{4/3} \cdot (\phi^*-\phi)^{-1}$. Thus they predict the same divergence of the viscosity with the particle concentration as that we measured although we have not observed the $\phi^{4/3}$-dependence.

In the particle concentration range that we consider here, our results are in close agreement with Stokesian Dynamics simulation of unstructured non-Brownian suspensions [Sierou and Brady (2001)]. Their results together with our experimental data are displayed in Fig. 12 where the inverse of the minimum reduced viscosity, $\eta_0/\eta_{min}$, is plotted versus the reduced particle volume fraction, $\phi/\phi^*$ (in our case, $\phi^*= 0.535$ and in the case of Stokesian Dynamics results, $\phi^*=0.64$). For relatively moderate solid volume fractions ($0.55<\phi/\phi^*<0.93$), the linear scaling is satisfactory for both results, even though the simulation results show a clear curvature. For higher particle concentrations, Sierou and Brady have proposed a different scaling of the viscosity with the reduced volume fraction that fits well to their simulations for this higher range of volume fractions. Unfortunately, we did not perform measurements at so high concentrations.

We note that in the work of Sierou and Brady (2001), the particles are homogeneously and isotropically distributed, whereas Kolli et al. consider the initial values of the shear and normal stresses after shear reversal, that correspond to a structured state where the non-hydrodynamic interactions have relaxed. As for our study, we consider the minimum of the viscosity that should be obtained for the less structured state of the sheared suspension, and our results compare well to the simulations of Sierou and Brady (2001). However, it is not clear to us whether the point is the isotropic structure or the lack of non-hydrodynamic interactions.

## VI. CONCLUSION

We have been able to characterize the bulk transient response of concentrated non-Brownian suspensions of various particle volume fractions after shear reversal, using a PIV technique. Our measurements confirm those previously obtained using classical viscosimetry [Gadala-Maria and Acrivos (1980), Kolli et al. (2002), Narumi et al. (2002)]: the transient viscosity mostly depends on the strain. After shear reversal, the viscosity undergoes a step-like reduction, decreases slower, passes through a minimum and increases again to reach a plateau value. The plateau viscosity obeys the classical Krieger-Dougherty law with an exponent n=2 and a packing concentration $\phi^*=0.535$ whose low value could be explained by interparticle forces. As for the minimum viscosity, its values compare well with the results of numerical simulation performed on random suspensions [Sierou and Brady (2001)] and is consistent with the predictions of Mills and Snabre (2009): in the range of particle volume fractions that we have studied ($0.55<\phi/\phi^*<0.93$), the minimum viscosity diverges as $(\phi^*-\phi)^{-1}$, suggesting that the suspension passes through an almost isotropic structure where the non-hydrodynamic forces have relaxed.

**ACKNOWLEDGMENTS**

This work was supported by the ANR (program ANR-08-BLAN-0048 - CSD 2)**.** We are grateful to L. Lobry and P. Mills for fruitful discussions.

**APPENDIX A: A POSSIBLE EXPLANATION FOR THE LOW VALUE OF THE PACKING FRACTION.**

Upon fitting the data of Fig. 3 with the Krieger-Dougherty law, the particle packing fraction, $\phi^*$, is found to be around 0.536. This value is much smaller than the usual values obtained for non-colloidal suspensions, $\phi_{nc}\approx 0.61$. We speculate that even though the particles are as large as 31µm in diameter, they undergo soft interactions. One can wonder about the nature of the inferred interactions. We can rule out the electrostatic interactions since we have measured the same value of $\phi^*$ when the particles were dispersed in an aqueous solution whose polarity is of course very different from that of the mineral oil used in the present experiment. The Van der Waals forces should also be discarded as their interaction range is too short for explaining the low value of the packing volume fraction. Furthermore, they should be weak since the refractive indices of the particles and of the liquid are matched. Finally, polymeric forces remain the only possible interaction between particles. We have to mention that we have attempted to wash the particles with deionized water and with ethanol without noting any change on the suspension rheology. According to the constructor (Microbeads AS), it is plausible that the particles are covered by polymer chains (cellulose) that served as surfactant during the manufacturing of the particles.

We can estimate the characteristic interaction length, $\delta$, by considering the excluded volume arising from the repulsion between particles:

$$\left(\frac{a+\delta}{a}\right)^3 \approx \frac{\phi_{nc}}{\phi^*} = \frac{0.61}{0.535} \tag{A 1}$$

where a is the particle radius. Eq.(A1) leads to $\delta \approx 1\mu m$.

Moreover, the presence of repulsive forces between particles is corroborated by the observation of an elastic behavior of the suspension which is illustrated in Fig. 13. This recovery test has been carried out on a $\phi=0.444$ suspension. The recovery curve (solid line) is almost well fitted by an exponential function (dashed line) with a Maxwell time, $\tau$, of about 2s which corresponds to an elastic modulus of approximately $\eta/\tau \approx 70$ Pa. Furthermore, we speculate that the repulsive forces between particles are responsible for a perturbation of the shear-induced microstructure. This has been observed during transient tests where shear stress steps have been applied always in the same direction after rest periods of different durations. The protocol is represented in Fig. 14.a and the bob angular velocity response is plotted in Fig. 14.b. The longer is the rest period, the higher is the angular velocity overshoot (or lower the viscosity). This would mean that, during shear periods, the particles are forced to approach each other in the compression direction and, when the shear is stopped, the particles move away one from each other under the repulsive interactions. Thus after a sufficiently long rest time, the shear-induced structure of the suspension tends to disappear and its viscosity approaches the minimum viscosity recorded during shear reversal. Again, this elastic behavior together with the small value of ϕ* suggest that some polymer molecules at the surface of the particles give rise to soft steric repulsion.

**APPENDIX B: COMPARISON WITH MACROSCOPIC NARROW-GAP MEASUREMENTS**

In this section, we perform classical rheometric measurements in a narrow-gap Couette geometry Z40Ti mounted on the rheometer Mars II (Haake, Thermofisher). The bob diameter is $2R_i=41.42$ mm, the gap is $g=R_o-R_i=0.99$ mm, its height is $L=55$mm. The transient viscosity of the $\phi= 47\%$ suspension is measured. After it had been poured in the cup, the suspension was shortly pre-sheared (total strain around 365) in order to obtain a well defined structure in the suspension. Then the shear reversal measurements were performed.

The raw viscosity measured for two values of the imposed shear stress (4.25 and 5.95 Pa) is displayed on Fig. 15, together with the corresponding wide-gap local measurements. The apparent viscosity is significantly lower in the narrow-gap cell, suggesting that wall slip occurs at the boundaries.

Jana et al (1995) performed local measurements in a narrow-gap geometry to determine the slip length at the wall. They used a Laser Doppler Velocimetry technique to measure the velocity profile across the gap. In this paper, they sheared the suspensions for 2 days before they acquired the velocity profile, so that they had to take the radial migration into account. The wall slip was characterized by an apparent slip velocity, i.e. the difference between the velocity of the walls and the velocity of the suspension at the same position. They showed that the slip velocity $u_s$ was related to the local shear rate $\dot{\gamma}$ and the local volume fraction $\phi$ by:

$$u_s = \frac{\eta_r}{q} \dot{\gamma} a \tag{B.1}$$

where a is the radius of the particles and $\eta_r$, the relative viscosity of the suspension. q is a constant that is determined experimentally: in [Jana et al (1995)], q=8.

In our work, the total strain during pre-shear is small, so that we can neglect migration. For comparison, Gadala-Maria and Acrivos (1980), in similar conditions, measured a characteristic strain for migration of around $4.32 \cdot 10^4$. Following Jana et al. (1995), we introduce a slip velocity at the bob $u_{si}$ and at the cup $u_{so}$. In the case of a purely viscous homogeneous liquid, the velocity profile is given by:

$$\frac{u}{r} = \frac{R_o^2 - r^2}{R_o^2 - R_i^2} \left(\frac{R_i}{r}\right)^2 \left(\Omega - \left(\frac{u_{si}}{R_i} - \frac{u_{so}}{R_o}\right)\right) + \frac{u_{so}}{R_o} \tag{B.2}$$

and the shear rate:

$$|\dot{\gamma}| = \left|r \frac{d}{dr}\left(\frac{u}{r}\right)\right| = \left(2 \frac{R_o^2}{R_o^2 - R_i^2}\left(\frac{R_i}{r}\right)^2\right)\left(\Omega - \left(\frac{u_{si}}{R_i} - \frac{u_{so}}{R_o}\right)\right) \tag{B.3}$$

where $\Omega$ is the bob angular velocity.

From (B.1), (B.2) and (B.3), the slip velocity is found at the boundaries, and the velocity profile can be computed.

$$\dot{\gamma}(R_i) = \frac{2R_o^2 \Omega}{R_o^2 - R_i^2} \frac{\frac{q}{\eta_r a}}{\frac{q}{\eta_r a} + \frac{2}{R_o^2 - R_i^2}\left(\frac{R_o^2}{R_i} + \frac{R_i^2}{R_o}\right)}$$

$$= \dot{\gamma}_{app} \frac{\frac{q}{\eta_r a}}{\frac{q}{\eta_r a} + \frac{2}{R_o^2 - R_i^2}\left(\frac{R_o^2}{R_i} + \frac{R_i^2}{R_o}\right)} \tag{B.4}$$

with $\dot{\gamma}_{app}$ the apparent shear rate, at the bob.

Using eq. (B.4) and setting that the stress is controlled:

$$\sigma = \eta_r \dot{\gamma} = \eta_r^{app} \dot{\gamma}^{app} \tag{B.5}$$

we obtain the expressions of the bulk shear rate, $\dot{\gamma}$, and relative viscosity, $\eta_r$, versus their apparent values $\dot{\gamma}^{app}$ and $\eta_r^{app}$:

$$\dot{\gamma} = \dot{\gamma}^{app} \left(1 - \frac{2\eta_r^{app}}{q} \frac{a}{R_o^2 - R_i^2} \left(\frac{R_o^2}{R_i} + \frac{R_i^2}{R_o}\right)\right)$$

$$\eta_r = \frac{\eta_r^{app}}{1 - \frac{2\eta_r^{app}}{q} \frac{a}{R_o^2 - R_i^2} \left(\frac{R_o^2}{R_i} + \frac{R_i^2}{R_o}\right)} \quad (B.6)$$

as well as the expression of the bulk strain:

$$\gamma = \int_0^{\gamma^{app}} \left(1 - \frac{2\eta_r^{app}(u)}{q} \frac{a}{R_o^2 - R_i^2} \left(\frac{R_o^2}{R_i} + \frac{R_i^2}{R_o}\right)\right) du \quad (B.7)$$

Figure (16) displays the PIV measurements together with the viscosity measured in the narrow-gap Couette rheometer after slip correction. The wall slip-corrected viscosity has been computed with q=6.8 that is the optimal value to match the plateau viscosity measured using PIV. If, like Jana et al., we take q=8, the plateau viscosity values are slightly smaller ($\eta_{plateau}$=56 Pa.s. for $\sigma$=4.25 Pa and $\eta_{plateau}$=54 Pa.s. for $\sigma$=5.95 Pa) but it is reasonable to expect that the q value depends on the particles and on the wall roughness and the wall-slip correction proposed by Jana et al. reconciles the narrow-gap and the PIV measurements at the plateau. However, the beginnings of the transients are quite different, suggesting that the slip velocity (B.1) is only valid when the sheared suspension has reached its steady state. According to figure 16, during the transient, eq. (B.1) underestimates the slip velocity.

At last, we would like to mention a striking result shown on Fig. 17 where the slip correction has been calculated upon replacing $\eta_r(\gamma)$ with its plateau value ($\eta_r(\gamma \to \infty)$), in eq. (B.1). Although we do not have any explanation for it, it appears clearly that the agreement is far better when eq. (B.1) is solved with $\eta_r(\gamma \to \infty)$ rather than with $\eta_r(\gamma)$.

**FIGURE CAPTIONS**

Figure 1: Flow curve of a $\phi=0.47$ suspension obtained in a narrow-gap Couette rheometer. Inset: power-law fit: $\ln(\sigma(Pa))=0.88.\ln(\dot{\gamma}(s^{-1}))+3.7$.

Figure 2: Experimental set-up.

Figure 3: Steady relative viscosity of the suspension versus the particle volume fraction. These measurements have been performed in the wide-gap Couette cell for an average shear rate of $0.02$ s$^{-1}$.

Figure 4: Typical image of a $\phi=0.444$ suspension recorded 14 mm above the cup bottom.

Figure 5: Long time evolution of the velocity profile in a $\phi=0.444$ suspension sheared in a wide-gap Couette cell (inner radius 14 mm and outer radius 24 mm). The bob spin rate is set to 1 rpm.

Figure 6: Long time evolution of the normalized viscosity profile for the $\phi=0.444$ suspension.

Figure 7: Transient velocity profiles after shear-reversal (at t=0 the applied torque passes from -250µNm to +250µNm) at different times (0.25, 1.25, 4 and 37.5 s). The dashed line represents the Newtonian profile. $\phi=0.444$.

Figure 8: Transient reduced viscosity after shear reversal versus local strain measured at $r/R_{out}=0.77$. $\phi=0.444$

Figure 9: Transient reduced viscosity after shear reversal versus local strain. The experiment has been performed for different applied torques that correspond to different shear stresses at the bob ($\sigma$=0.85, 2.12, 4.25 and 5.95 Pa). $\phi$=0.47.

Figure 10: Transient viscosity after shear reversal for different particle volume fractions vs local strain. $\langle\dot{\gamma}\rangle_{plateau} \approx 0.1$ s$^{-1}$. The applied torque values have been chosen in order the bob angular spin rate to be roughly 0.5 rpm at the plateau whatever the particle volume fraction and thus the viscosity.

Figure 11: Scaling law for the plateau viscosity and the minimum viscosity. The applied torque values have been chosen in order the bob angular spin rate to be roughly 0.5 rpm whatever the particle volume fraction and thus the viscosity.

Figure 12: Inverse of the relative viscosity $\eta_0/\eta$ versus $\phi/\phi^*$: (●) minimum viscosity from PIV measurement. (■) High-frequency viscosity from Stokesian dynamics (Sierou and Brady (2001)].

Figure 13: Recovery test after shear stress cancellation. $\phi$=0.444.

Figure 14: Sequence of shearing and shear stress break with increasing of the break duration. a) Applied torque. b) Bob angular velocity response. The velocity overshoot amplitude increases with the break duration. $\phi$=0.444.

Figure 15: Viscosity response to a shear stress reversal for a ϕ=0.47 suspension. Wide-gap + PIV measurements: σ=4.25 Pa (▽), σ=5.95 Pa (◇). Narrow-gap measurements: σ=4.25 Pa (──), σ=5.95 Pa (- - - -).

Figure 16: Viscosity response to a shear stress reversal for a ϕ=0.47 suspension. Wide-gap + PIV measurements: σ=4.25 Pa (▽), σ=5.95 Pa (◇). Narrow-gap measurements corrected for the wall slip: σ=4.25 Pa (──), σ=5.95 Pa (- - - -).

Figure 17: Viscosity response to a shear stress reversal for a ϕ=0.47 suspension. Wide-gap + PIV measurements: σ=4.25 Pa (▽), σ=5.95 Pa (◇). Narrow-gap measurements corrected for the wall slip: σ=4.25 Pa (──), σ=5.95 Pa (- - - -). The wall slip correction is calculated using the plateau viscosity in eq. (A.2.1).

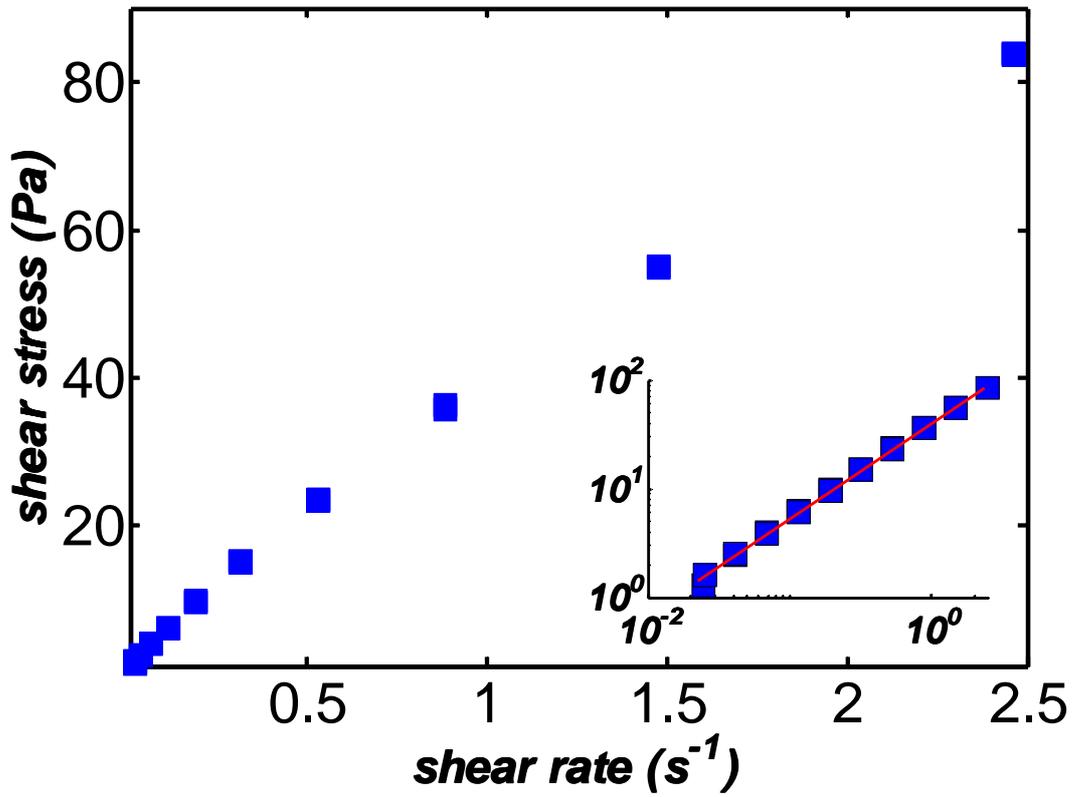

Figure 1

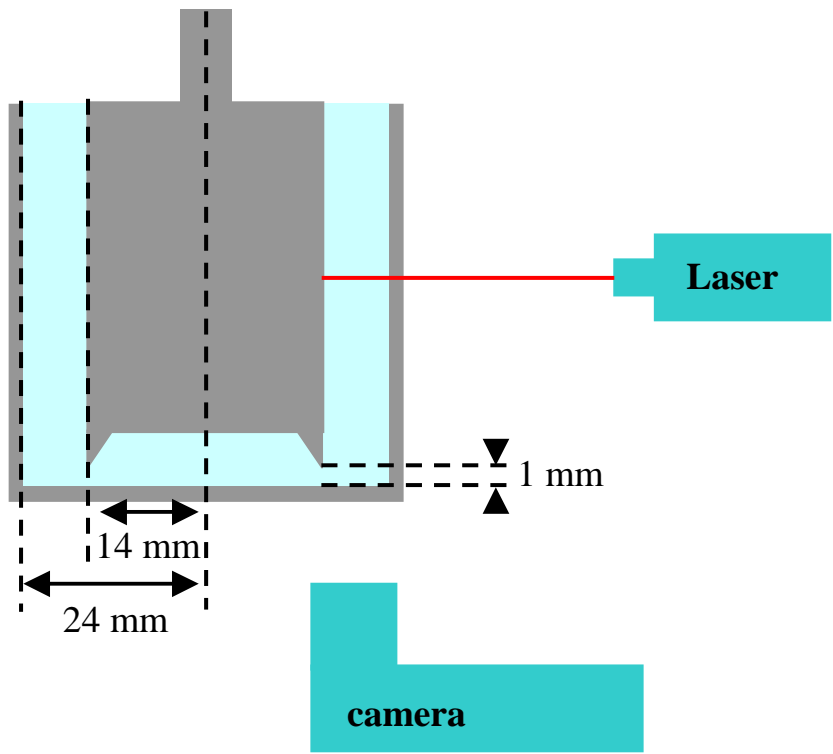

Figure 2

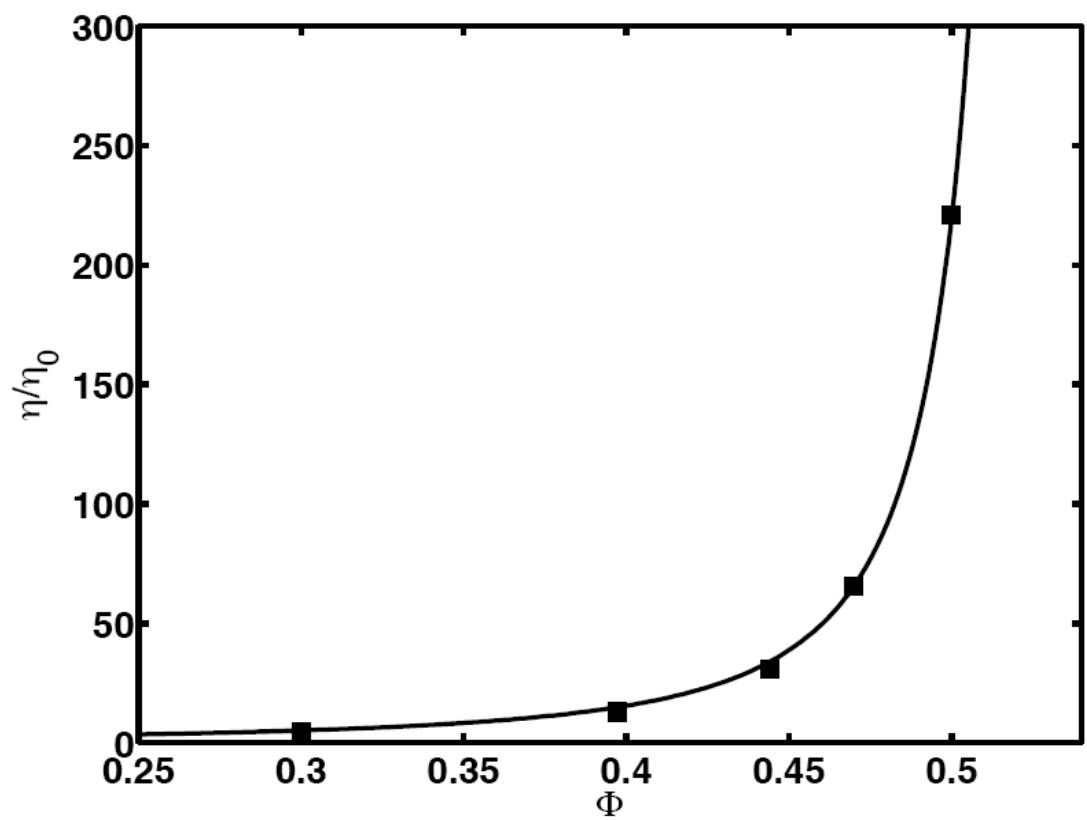

Figure 3

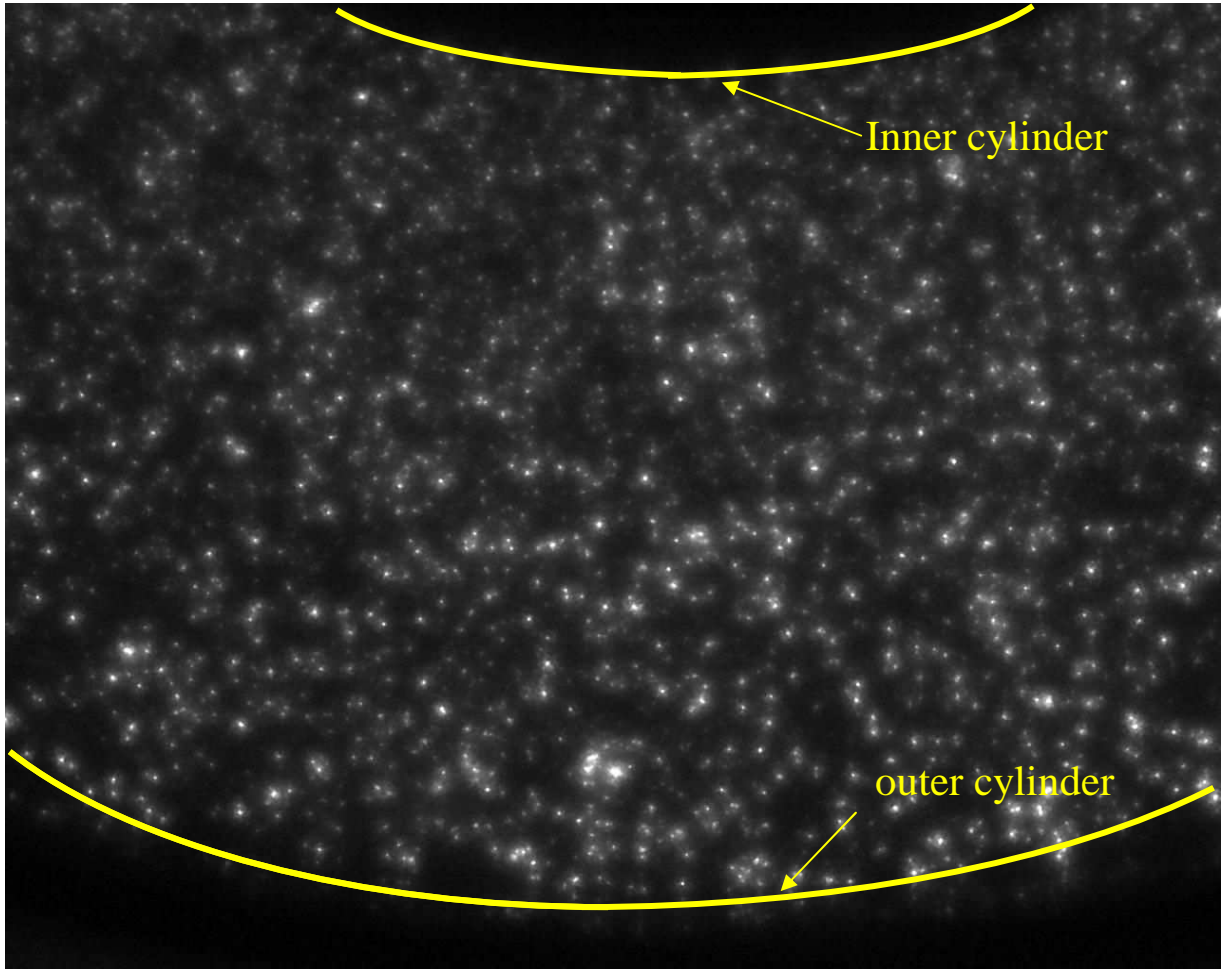

Figure 4

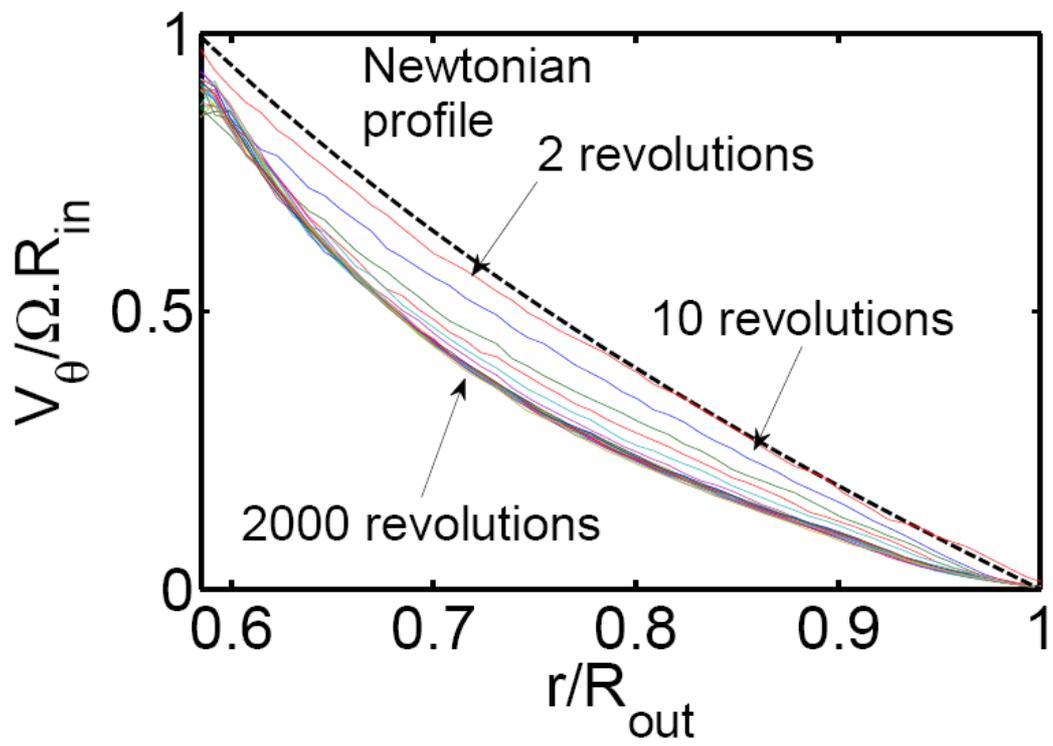

Figure 5

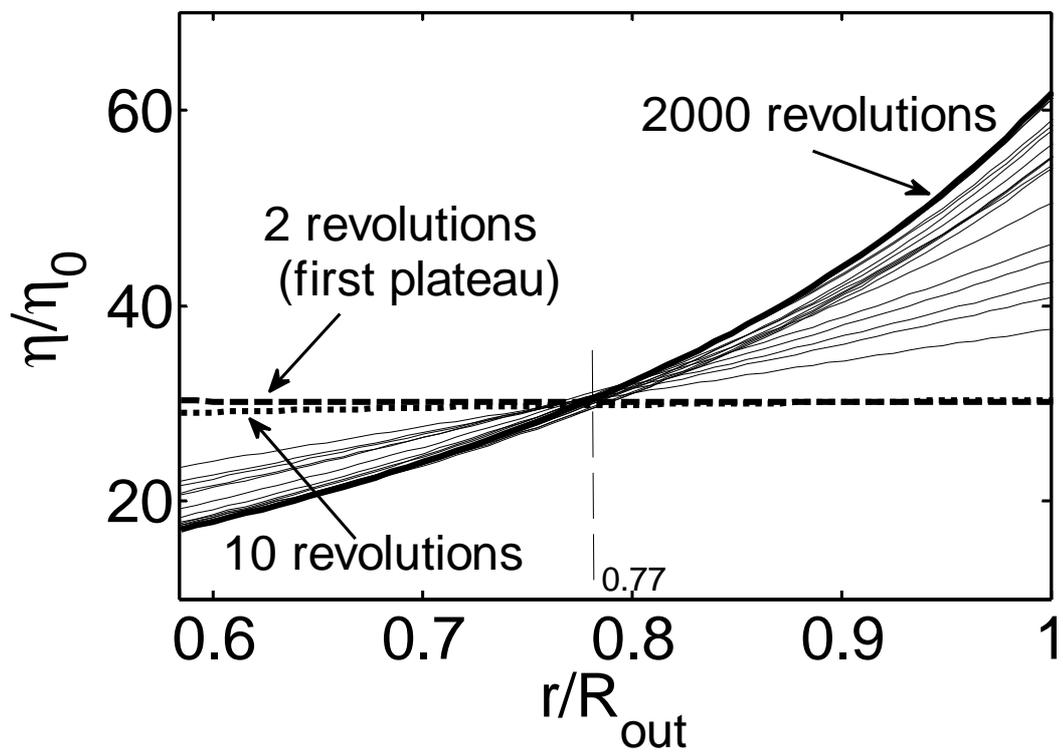

Figure 6

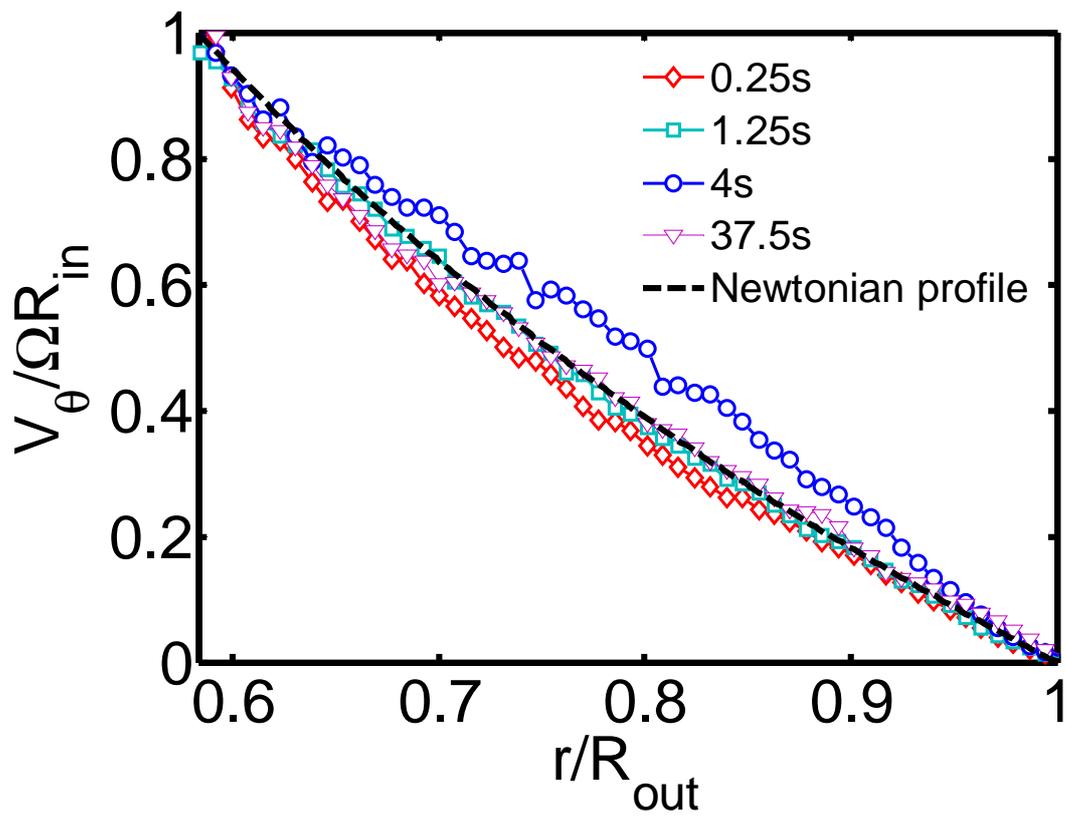

Figure 7

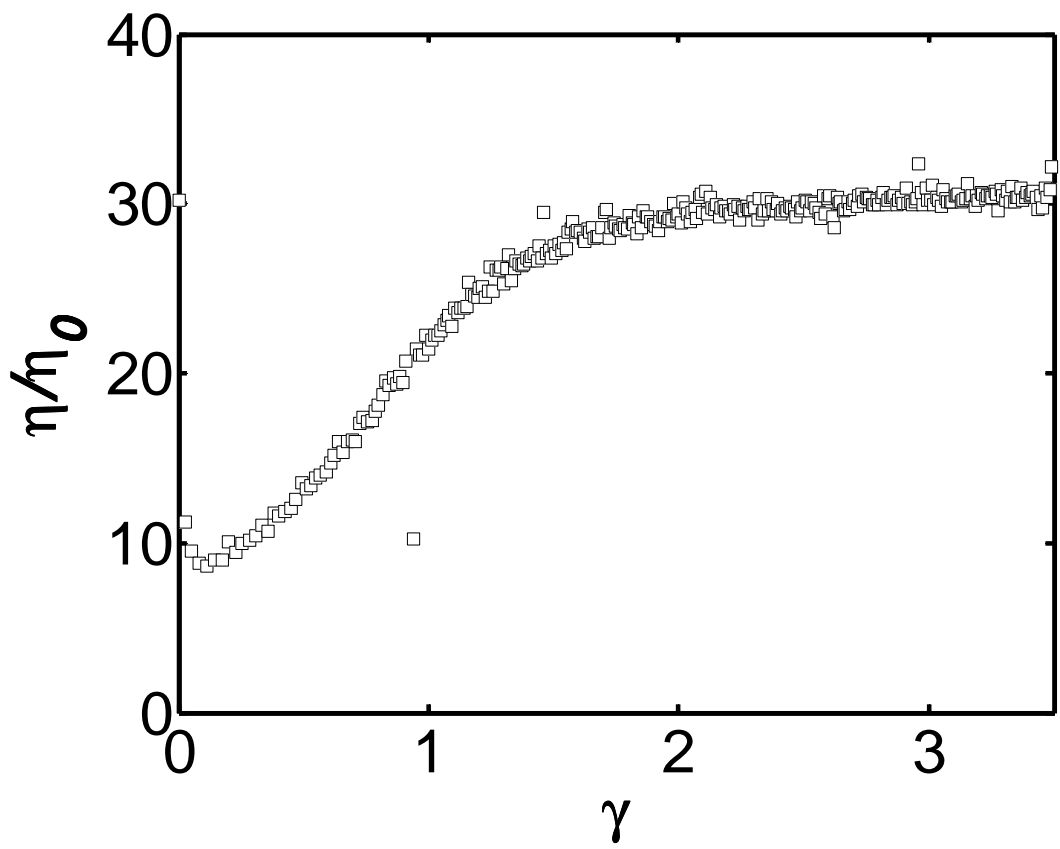

Figure 8

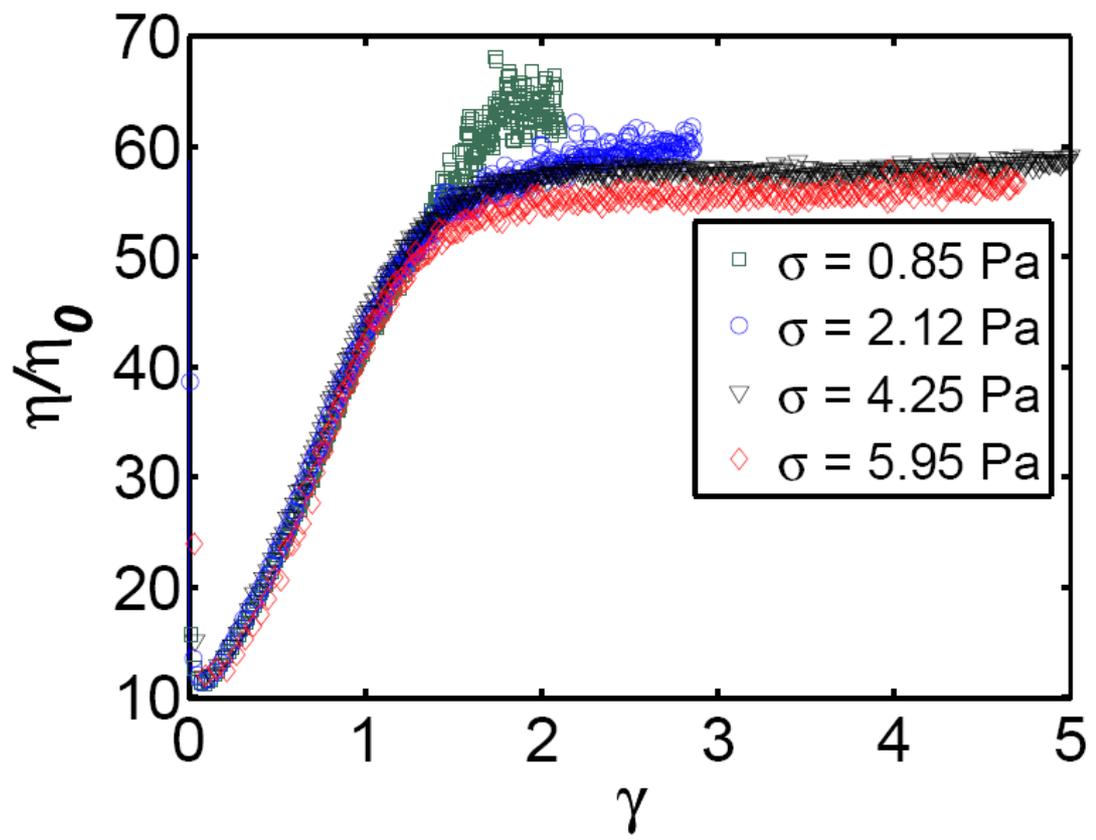

Figure 9

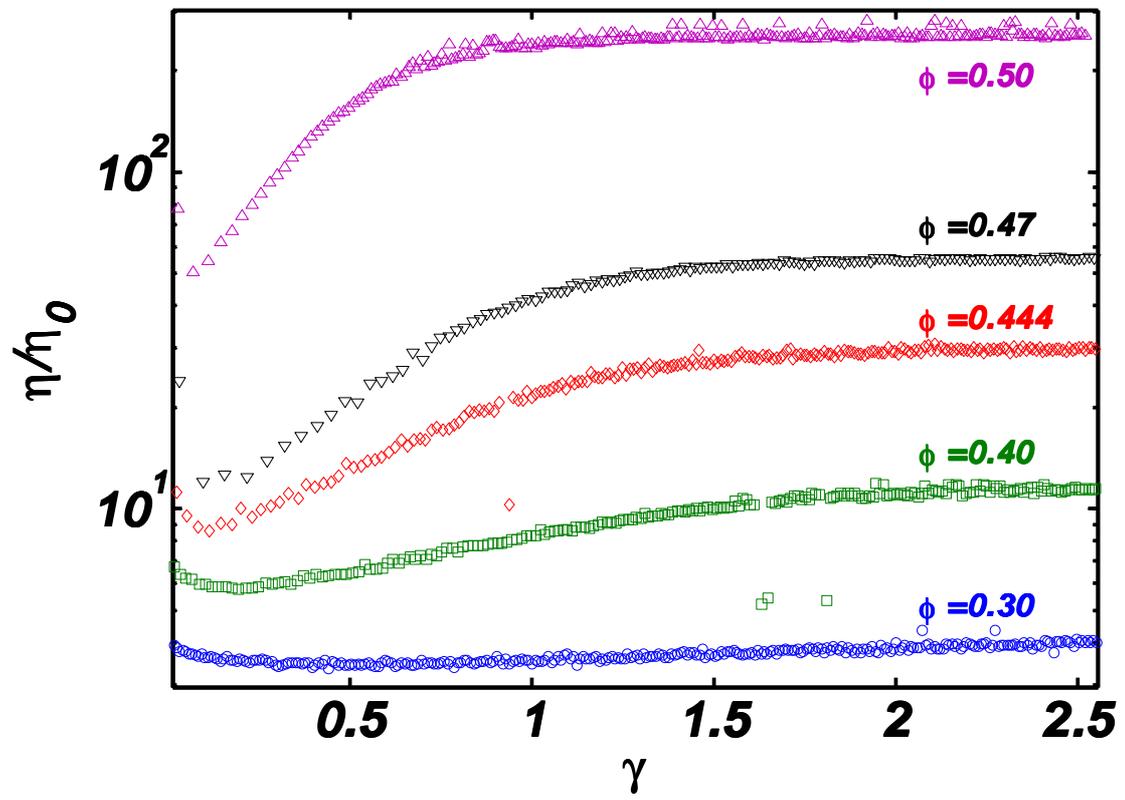

Figure 10

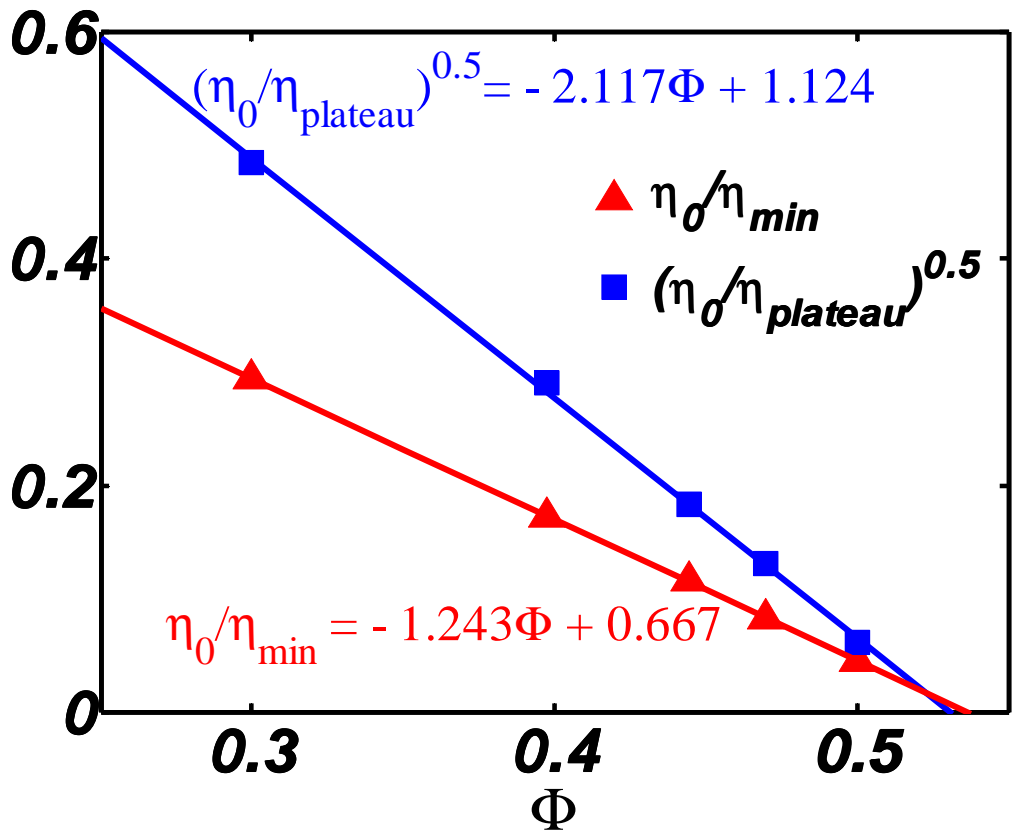

Figure 11

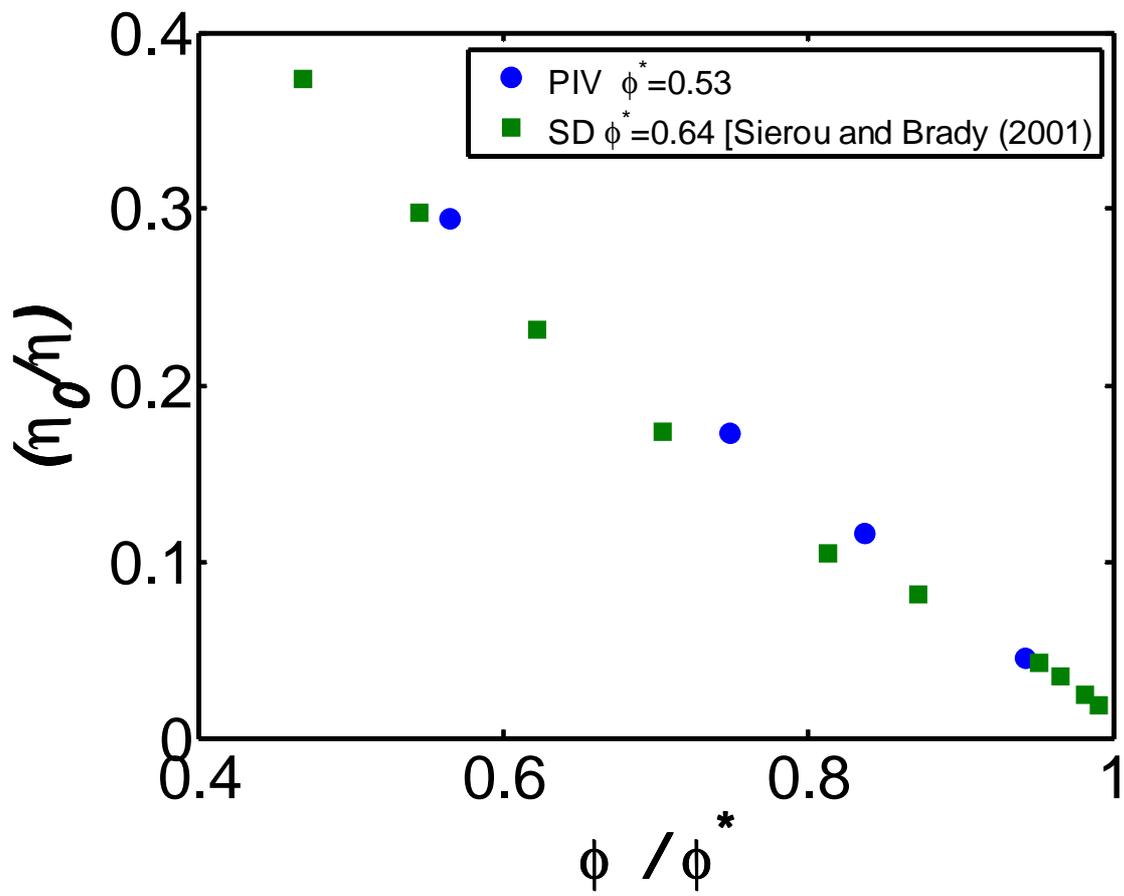

Figure 12

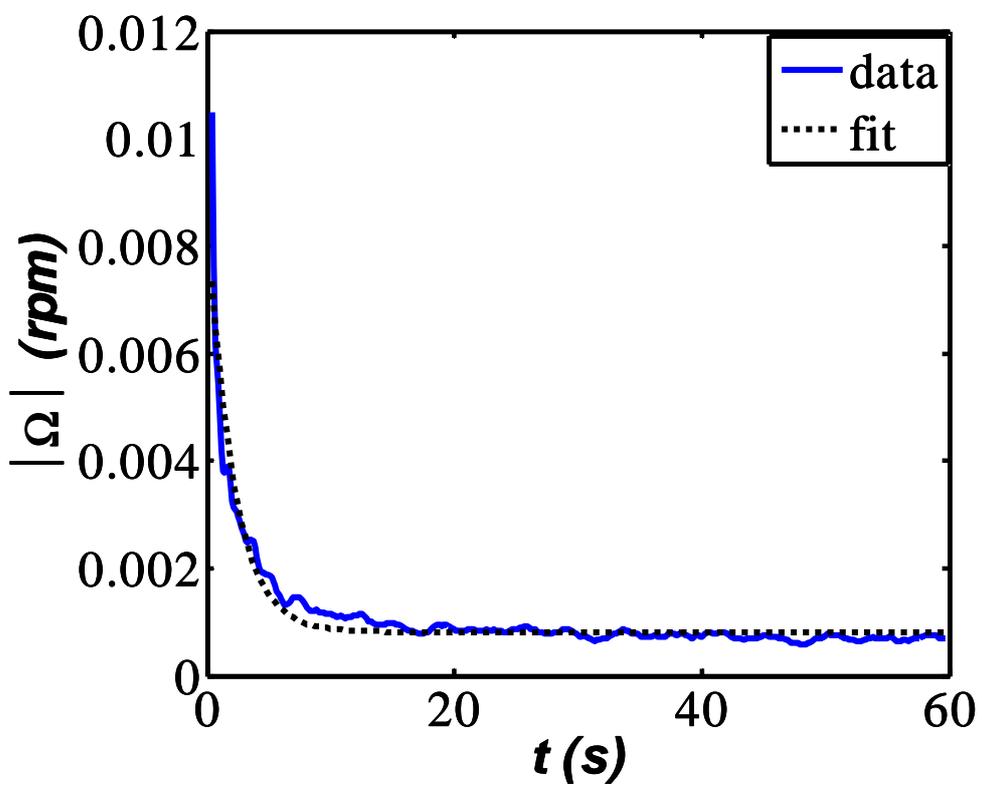

Figure 13

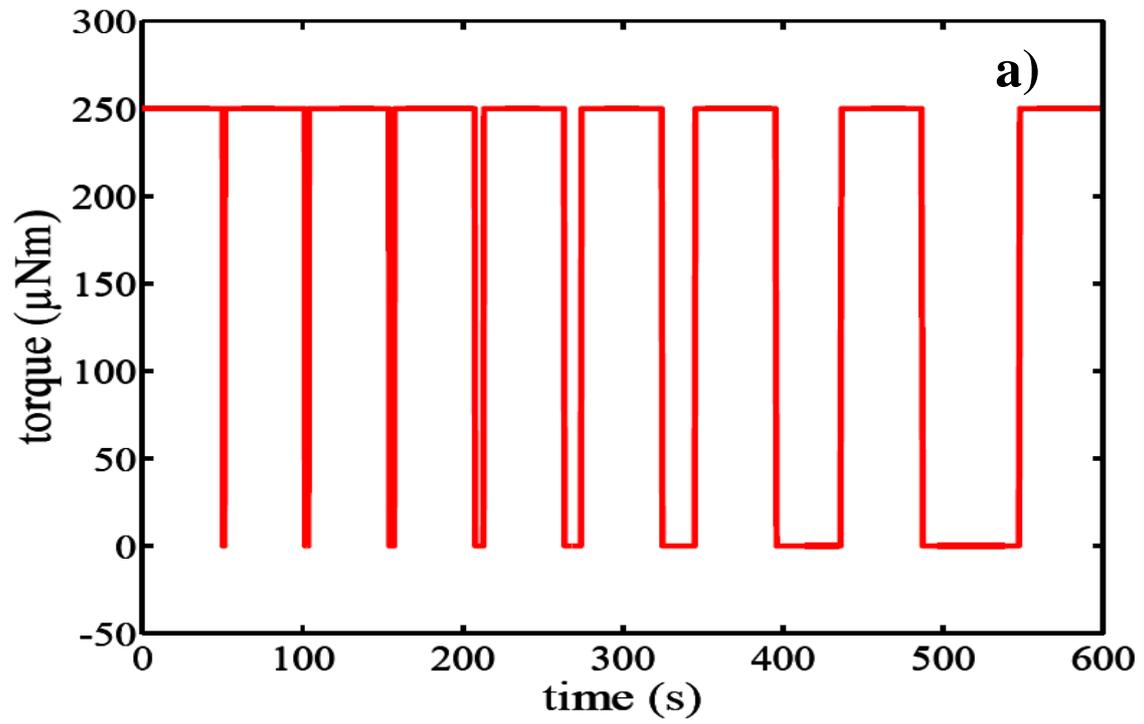

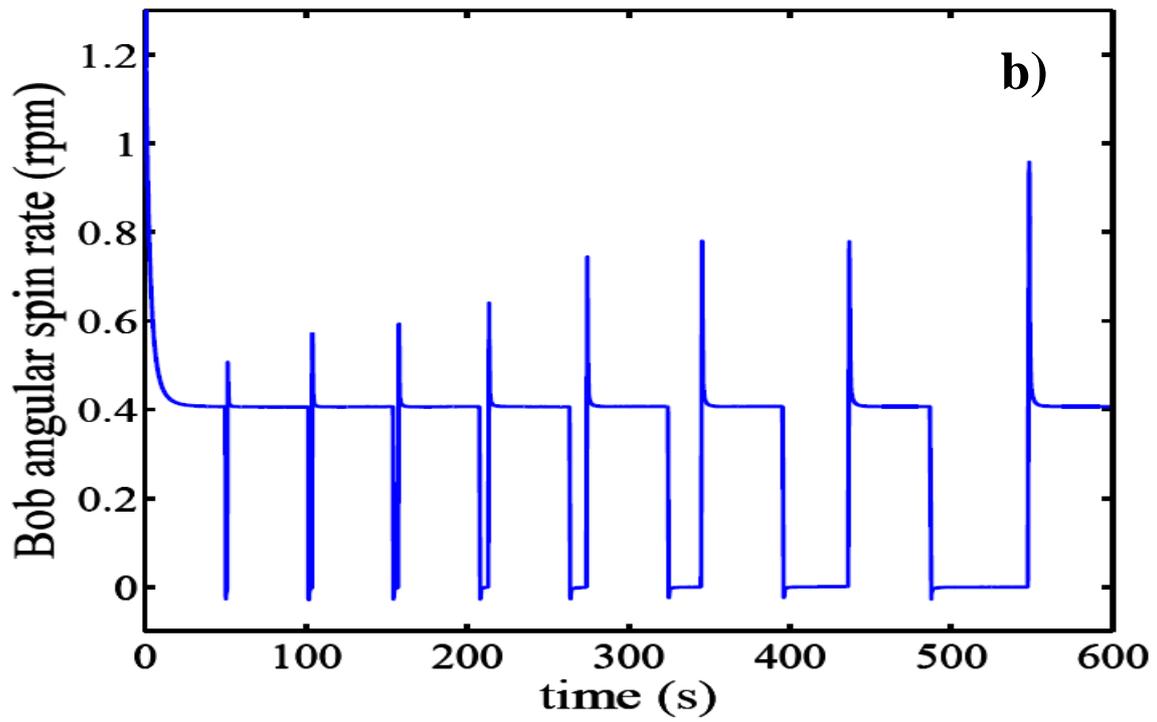

Figure 14

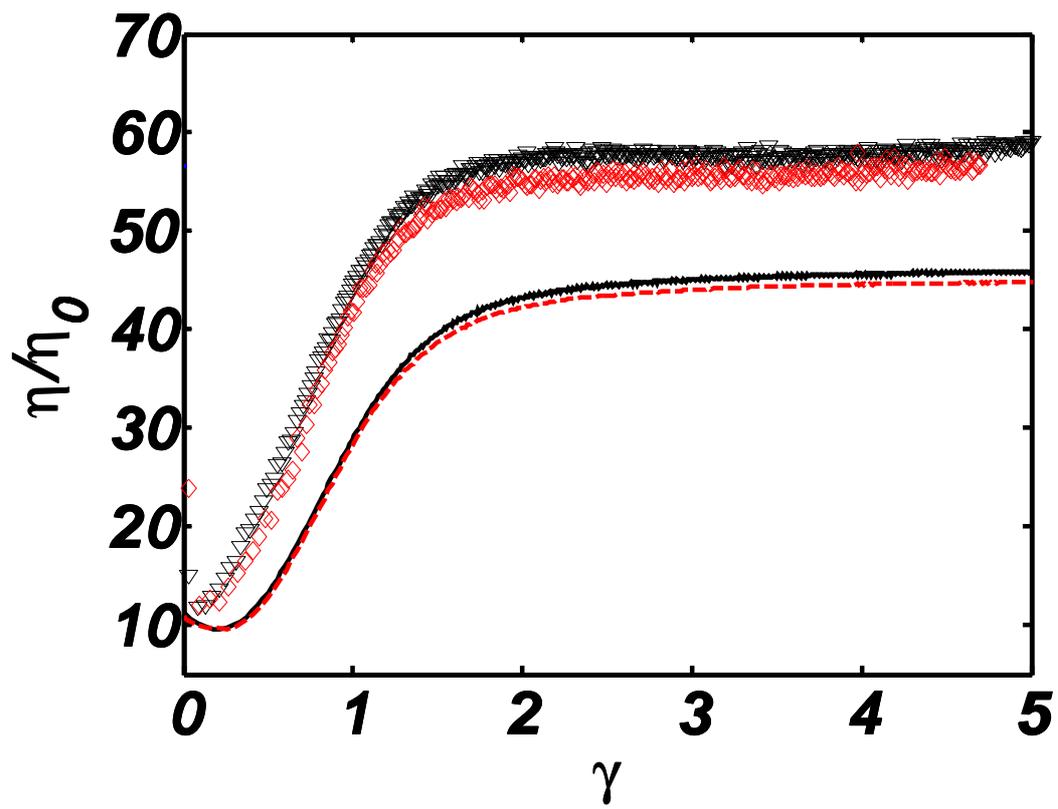

Figure 15

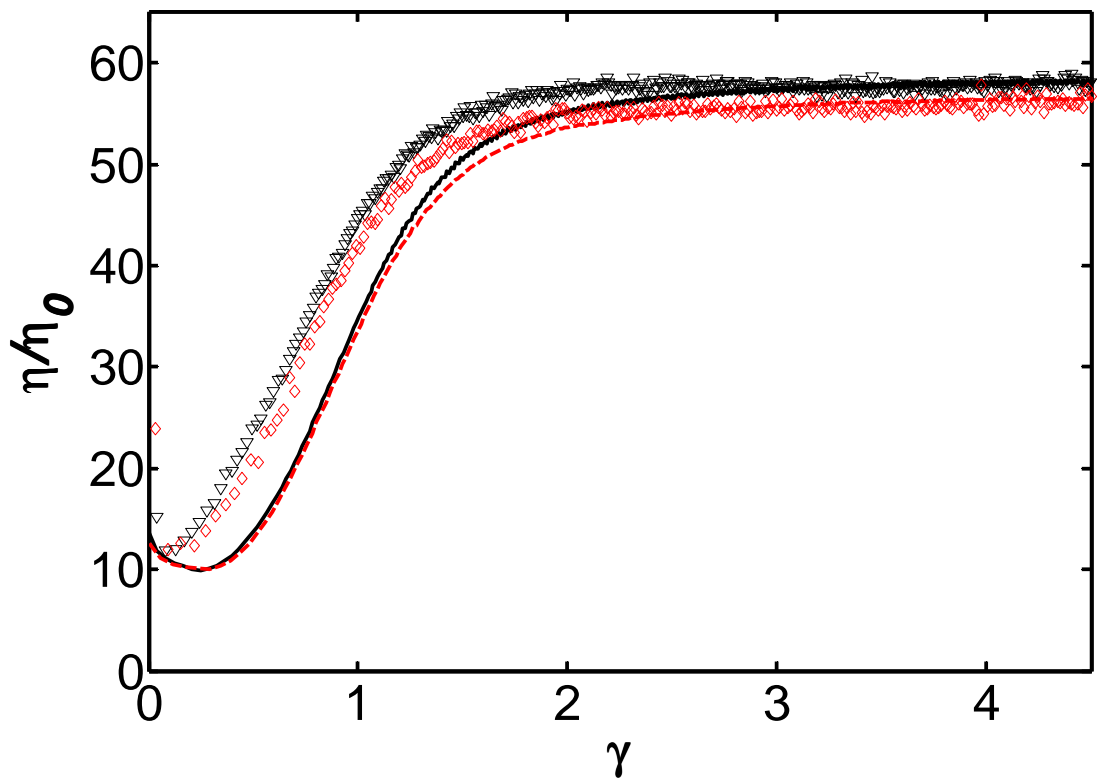

Figure 16

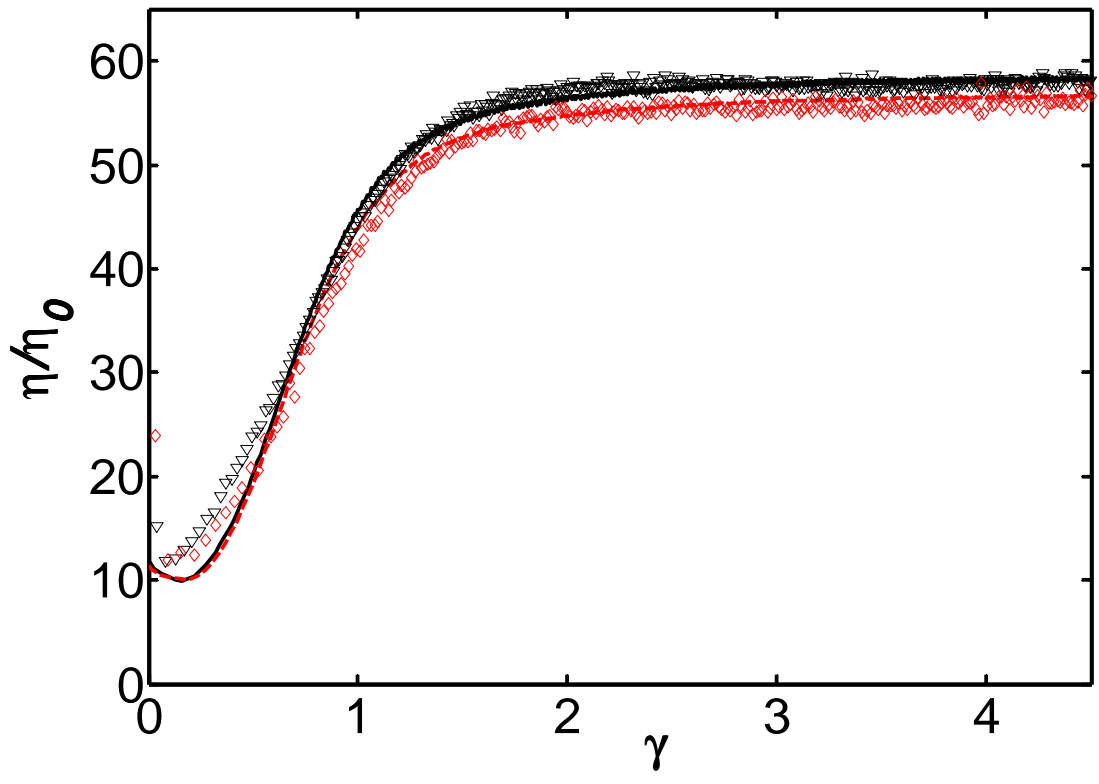

Figure 17